\documentclass[10pt,a4paper,reqno,twoside,psamsfonts,titlepage]{amsart}
\usepackage[latin1]{inputenc} \usepackage[T1]{fontenc}
\usepackage{textcomp} \usepackage[british,german]{babel}
\usepackage{a4wide}
\usepackage{newcent} 
\providecommand{\href}[2]{#2} 
\providecommand{\hypersetup}[1]{}
\newif\ifpdf \ifx\pdfoutput\undefined
 \pdffalse 
 \else
 \pdfoutput=1 
 \pdftrue \fi
\ifpdf 
  \usepackage[pdftex]{graphicx} 
  \usepackage{hyperref}  
  \pdfcompresslevel=9 \hypersetup{pdfauthor= {Andreas U. Schmidt}}
  \else
  \usepackage[dvips]{graphicx} 
 \usepackage[dvips]{hyperref}%
\fi
\begin{document}
\selectlanguage{british}
\pagestyle{empty}
%
%
\title{Security for Distributed Web-Applications via Aspect-Oriented Programming}
\author[N.\ Kuntze]{Nicolai Kuntze} 
\author[Th.\ Rauch]{Thomas Rauch} 
\author[A.\ U.\ Schmidt]{Andreas U.\ Schmidt${}^\dag$} 
\thanks{\textit{Address:} Fraunhofer--Institute for Secure Information Technology SIT,
Dolivostraße 15, 64293 Darmstadt, Germany}
\thanks{\textit{Tel.:} +49-6151-869-60227, \textit{Fax:} +49-6151-869-704}
\thanks{\textit{E-mail address:}  \href{mailto:{Nicolai.Kuntze,Thomas.Rauch,Andreas.U.Schmidt}@sit.fraunhofer.de}{\{Nicolai.Kuntze,Thomas.Rauch,Andreas.U.Schmidt\}@sit.fraunhofer.de}}
\thanks{\textit{URL address:}\href{http://www.math.uni-frankfurt.de/~aschmidt}{http://www.math.uni-frankfurt.de/\~{}aschmidt}}
\thanks{${}^\dag$Corresponding author.}
\begin{abstract}
  Identity Management is becoming more and more important in business
  systems as they are opened for third parties including trading
  partners, consumers and suppliers. This paper presents an approach
  securing a system without any knowledge of the system source code.
  The security module adds to the existing system authentication and
  authorisation based on aspect oriented programming and the liberty
  alliance framework, an upcoming industrie standard providing single
  sign on. In an initial training phase the module is adapted to the
  application which is to be secured.  Moreover the use of hardware
  tokens and proactive computing is demonstrated. The high
  modularisation is achived through use of AspectJ, a programming
  language extension of Java.
\end{abstract}

\maketitle
%
%
\section{Introduction}
Outsourcing of IT-infrastructure has become a common practise for
companies and administrations.  Pressure toward rationalisation has
spurred desires to relocate central parts of business processes to
external service providers.  This becomes critical from a security
viewpoint, in particular if employees of
partner companies and field staff are to be given restricted yet
direct access to a company's data
stock within complex workflows~\cite{Khalf04,DACH04-Mehrhoff}.  
For instance, the Distributed Management Task
Force has developed a comprehensive set of standards aiming at
implementing security in distributed IT Systems, and Sun Microsystems has
released a toolkit for Web-based Enterprise Management~\cite{WBEM}
based on them.  However, to deploy security functionality for a total
system based on an explicit security policy with this approach, a
considerable effort has to be spent, in particular if legacy systems
need to be incorporated \cite{DACH04-MakoSi}. Especially  for
small and medium-sized enterprises, this is not compatible with the pursuit of
efficiency in, and through, IT-outsourcing.

We report on an approach, and its prototypical implementation, with
which an existing web-application can be augmented by the three
fundamental security functionalities authentication, authorisation,
and accounting (AAA).  Particular traits of the method presented are
its universality, flexibility, and scalability.  The conceptually new
combination of common, freely available technologies with the paradigm
of \textit{aspect-oriented programming}~\cite{AOPBible} presupposes
only a minimal knowledge of the system to be secured, and in
particular its source code.
\begin{figure}[htbp]
  \centering \ifpdf

  \rotatebox{-90}{\resizebox{0.6\textwidth}{!}{\includegraphics{uebersicht.png}}}
  \else
    \rotatebox{-90}{\resizebox{0.6\textwidth}{!}{\includegraphics{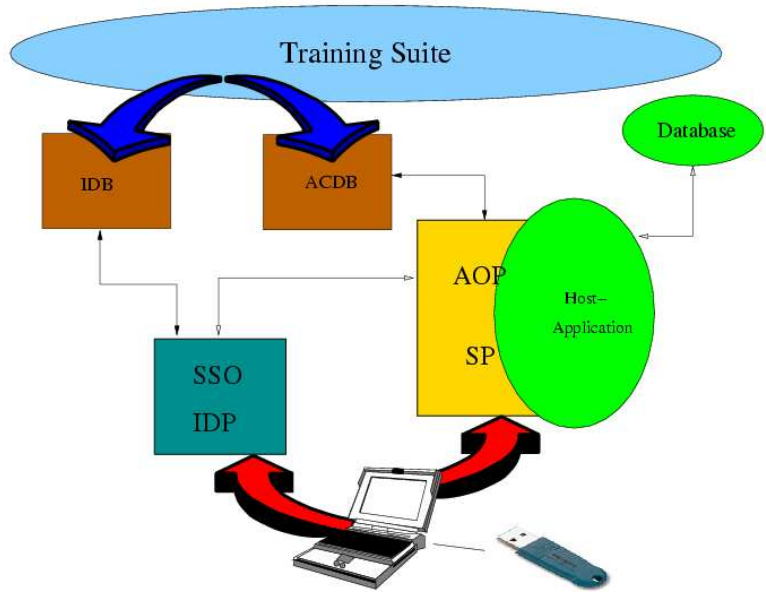}}}
\fi
  \caption{Aspect-oriented security architecture.}
  \label{fig:Architektur}
\end{figure}

The sketch above shows the architecture.  The application hosted by the
service provider SP is joined at decisive points with a generic
security module, the \textit{security aspect} using the AspectJ
framework, which thus gains control in all situations in which AAA
functionality becomes necessary.  The security aspect is the central
interface to an identity management system, realised using the
reference implementation SourceID of the Liberty protocol
suite~\cite{libertyTutorial}.  The latter single-sign-on (SSO) and
accounting system of the identity-provider IDP exerts all tasks for
authorisation and user management, providing the basis for the
role-based access control system~\cite{RBAC}.
The RBAC model is implemented by the
security aspect, which grants clients access to the
application.

The data bases for accounting (IDB) and access control (ACDB) are
built up by an intuitive method through an interactive process.  The
productivity phase of the system is preceded by a training phase
during which all relevant workflows, i.e., sequences on web pages and
contained entry masks, are learnt.  Recorded state transitions in
workflows are associated with authorised roles, allowing also for
ancillary conditions, in particular states of the host application's
data base. The security model follows the paradigm that only the
recorded workflows are allowed and any other state transition is
prohibited.

The security system is also flexible with respect to authentication
methods.  As an example, the combination of user-name/password
authentication with authentication through ownership, here a hardware
token~\cite{wibukey}, is demonstrated.  The security system can be
linked with arbitrary authentication methods, which in turn can be
associated to roles in arbitrary combination.  In particular, the
usage of an intelligent hardware token enables proactive methods for
security enhancement on the client's side, by guaranteeing freshness
of the authentication by periodic polling for the token's presence.

The paper is structured so as to provide some background on the used methods
first and then to detail the system's architecture, implementation  and deployment.
Readers mainly interested in the latter aspects might want to skip
to Section~\ref{arch}.
Section~\ref{liberty} gives an overview of the authentication system
based on the Liberty Alliance Protocol.  The proactivity
Section~\ref{proactivity} describes the usage of a hardware-token and
the proactive component.  Section~\ref{aspect} illustrates the usage
of the Aspect Oriented Programming Language and the functioning of the
security aspect.  Section~\ref{rbac} goes into details about RBAC and the
interdependencies between user, roles, and workflows.  The
architecture Section~\ref{arch} describes the system architecture and
the interaction of the system's modules.  Chapter~\ref{proto} provides
a description of the prototypical implementation and gives an example
for a typical deployment process.  The conclusions discuss pros and
cons of the present approach.
\section{Identity Management}\label{liberty}
Single sign-on (SSO) systems become more and more attractive in
particular because they provide a way to ease the handling of multiple
applications and accounts for users. In a closed infrastructure, SSO
enables a centralised user management allowing or denying the user
access to company resources. Different SSO systems, like Kerberos, have been developed
over the last years.

Opening an IT-infrastructure toward the Internet creates new business
concepts and therefore new requirements. In this context, an \textit{identity}
is a name with corresponding attributes. Web applications are
performing on behalf of things with identities like computers, humans,
or companies. Due to this, all Web applications have common  requirements
for security and related functionalities: 
authentication, authorisation, auditing, integrity,
confidentiality, non-repudiation, and trust. Identity Management helps
to implement these demands by \textit{federation} of identities.
In \cite{IndetityPrimer} five use cases for identity federation are
identified. The {\bf Internal Web Single Sign-On} covers log-on at the
operating system. The second use case {\bf External Web Single
  Sign-On} enables a user in company A to use resources of company B.
The following use cases extend  this core concept.
{\bf Attribute Exchange} enables company B to make a separate request
for additional user attributes.  By this, the user
interaction can be personalised and thus become more convenient. Creation and
deletion of accounts in remote security domains is described by {\bf
  Federated Provisioning}. In the fifth use case, {\bf Federated Web
  Services} are covered.

\begin{figure}[htbp]
   \centering \ifpdf

   \resizebox{0.6\textwidth}{!}{\includegraphics[width=0.5\textwidth]{Liberty_arch.png}}
   \else
\resizebox{0.6\textwidth}{!}{\includegraphics[width=0.5\textwidth]{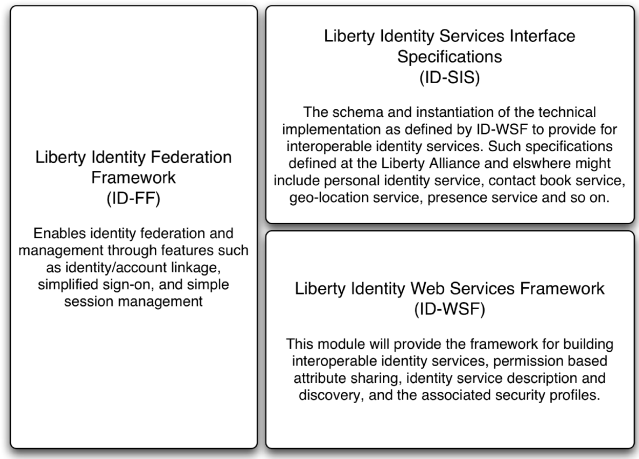}}
\fi
  \caption{Liberty Alliance Architecture}
  \label{fig:libertyarch}
\end{figure}

The Liberty Alliance~\cite{liberty}, which represents a broad spectrum
of associations and companies, tries to fulfil these demands through
a set of protocols (see Figure \ref{fig:libertyarch})
covering most aspects of federated identities, like opt-in
account linking (a synonym for ID federation proper), 
simplified sign-on (SSO), basic session
management, pseudonymity, and metadata (policy) exchange.
The goal is to provide a common platform for federated user
authentication systems.  For instance, an airline and a car rental
company can cooperate using ID-FF, providing the customer with
combined offers and the possibility for only one login procedure for
both companies' websites.
A general disclaimer applies with respect to identity federation:
Trust is more than a pure technical problem. Before federated identity
networks can be established there must be working agreements defining
the scope of a co-operation and commonly agreed policies.

\begin{figure}[htbp]
   \centering \ifpdf
   \resizebox{0.5\textwidth}{!}{\includegraphics{circleoftrust.png}}
   \else
\resizebox{0.5\textwidth}{!}{\includegraphics{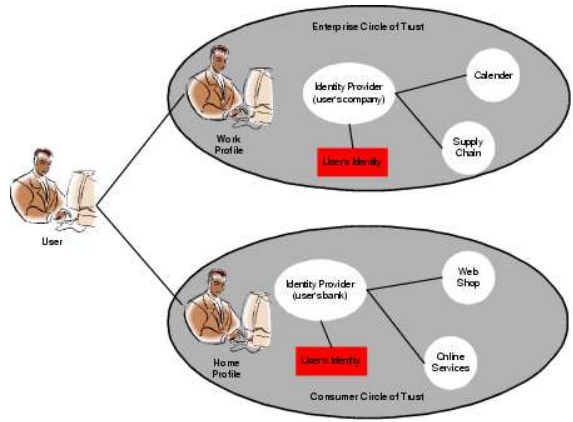}}
\fi
  \caption{Circle of Trust}
  \label{fig:circleoftrust}
\end{figure}

The Liberty framework is based on the assumption that nearly all on-line
applications are equipped with a user management system to which the
user has to present his credentials upon login.  ID-FF provides a way
to reflect contractual relationships between different companies with respect
to the login procedures of their different websites. Such an on-line
relationship is called a Circle of Trust, see
Figure~\ref{fig:circleoftrust}, and contains an Identity Provider (IDP)
and different Service Providers (SP). The IDP is the trusted central
login portal which manages and authenticates
users.  The SP generally offers services to users who are in turn
authenticated by the IDP on an SP's request.  The cooperation of IDPs
and SPs is not limited and static, different IDPs and SPs can be
integrated on-the-fly.

In our prototype, we enable the integration of the ID-FF protocol into
an application without knowing and changing its source code. This is
possible due to the fact that the security aspect encapsulates the
complete application and integrates the service provider part of the
ID-FF protocol by using the SourceID framework~\cite{SourceID}.  As we
do not interfere with the internal execution of the host application
only the ID-FF protocol is used in the prototype.

The AAA-Architecture is subdivided into the
authentication part of the IDP and the authorisation and accounting
part of the security aspect encapsulating the application. This presupposes
a trust relationship between
the SPs and the IDP which in turn is realised using the ID-FF protocol by 
exchanging granting tickets issued by the IDP. Authentication is in the domain of
the IDP, audit and authorisation is located in the SP (see Section~\ref{arch}).
Within this scenario it is possible to model security and
implement special security and authentication methods, in particular
hardware token based authentication, at the IDP.  Furthermore administration is
facilitated by the centralised accounting database (IDB).

At the moment there are big changes in the landscape of federated
identity protocols.  The Liberty Alliance has signed a Memo of
Understanding with OASIS \cite{SAML} in October 2004 and will extend
their Inter-operable Testing Program to include SAML 2.0 \cite{SAML20}.
It seems that Liberty, as a protocol, is largely drying up and the
ID-FF 1.x protocol is the end of its standardisation life-cycle, in view of the
fact that its functionality has been wrapped into SAML 2.0.
\section{Proactivity}
\label{proactivity} 
Using federated identities in the context of the Internet needs (1) good
authentication methods and (2) a user awareness about the value of their
identity tokens. (1) is addressed by adding to the standard user
name/password scheme an authentication by hardware token.
Requirement (2) allows for various solution approaches. We chose to
educate users by proactive methods.

During initial authentication, the user has to present his credential,
user name and password.  To check for a user's continued presence,
periodical re-authentication is often used as an additional security
measure.  A periodical asking for the password is very bothersome and
users tend to avoid such security measures.  We use the USB hardware
token Wibu-Key as authentication credential and re-authentication
system.  For authentication, a hardware-token based approach has the
advantage that no user interaction is needed after the initial
authentication and that the host system's identity is secured.

As a definition~\cite{proactive}, a proactive system is
\begin{quotation}
  1) working on behalf of, or pro, the user, and\\
  2) acting on their own initiative.
\end{quotation} 
The applet is loaded into the users computer after login and checks if
the hardware token is present and whether the user
is working (keystroke and mouse-activity).  If the user is inactive
for a pre-defined period, the applet sends a message to the IDP and the
user is automatically logged off.  Thus, this applets assists the user if
he forgets to unplug the hardware-token when leaving the workplace.
The proactive applet is neutral with respect to authentication method
and the criteria defining users' inactivity, and the re-authentication
period is configurable.  Needless to say, besides being a genuine
security measure, this opens ways to protocol users' behaviour that
may threaten their privacy.
\section{Aspect Oriented Programming}\label{aspect}
In every object-oriented software design there are \textit{core concerns}.
For instance, in an robotic system these are motion
management and path computation. 
These concerns are located in one class and are not
needed in any other scope. Other concerns are common to many of a
system's modules, like logging, authorisation, and persistence. These
system-wide concerns are called \textit{crosscutting concerns} in the
aspect-oriented parlance and the
reimplementation of one issue in different modules is called
\textit{code scattering}.

Starting in the year 1997 with \cite{AOP_start1} and \cite{AOP_start2}
the paradigm of the aspect oriented programming (AOP) was developed.
AOP attempts to isolate crosscutting concerns in special modules called
aspects. According to \cite{AOPBible}, developing a system using AOP
involves three steps: (1) Aspectual decomposition, (2) Concern
implementation, and (3) Aspectual recomposition. In step (1) the
crosscutting and core concerns are identified and separated. Step (2)
comprises the implementation and testing of each concern independently. Finally in
step (3) recomposition rules are specified by creating aspects. The
process of recomposition is referred to as \textit{weaving}.

Weaving is the key feature of this programming paradigm. It uses
weaving rules to define the point of execution of an aspect in the program flow.
AOP distinguishes between where a concern has to be inserted into the
program and when the concern has to be executed. The `where' is defined
by the \textit{Pointcut} which is an identifiable point (join Point)
in the execution of a program,e.g., a method call. The
\textit{Advice} is the aspect code to be executed. In the body of the
Advice the execution time is defined by the keywords \textbf{before},
\textbf{after}, and \textbf{around}.

Weaving modifies the Java-bytecode. This means that it is possible to
change a program's behaviour by adding features or removing them, even
after the compilation process. The resulting program is native java
bytecode without any need for a special runtime environment. In
particular, all security checks performed by the java runtime
environment (JRE) are left unharmed. Thus, the JRE still guarantees
memory and type safety and restricts the usage of system resources.
 
The main achievement of AOP, as underlined in~\cite{AOSD}, is a better
code comprehensibility due to the higher modularisation. 
Aspect-orientation also entails a higher re-usability
of developed modules. Higher modularisation also means better
verifiability of the single modules.  On the
negative side, it is often very difficult to verify the focus of the
pointcut.  Further research is needed to ensure that all intended
methods are selected exclusively by them.

In general, authentication and authorisation are considered as good
examples for AOP based implementations In~\cite[chapter 10]{AOPBible},
a module using the Java Authentication and Authorisation Service
(JAAS) is shown as example. In the simplest approach both concerns are
executed before every execution of a method, checking the user
authentication and the authorisation to use this particular method.
This tends to slow down the application and thus the developer must
carefully decide in which points to use the Pointcut destining the
security checks.

AOP opens further possibilities for software development through
`Policy Enforcement'. Compile time declarations are used to
formalise rules in the development of a system. For example a very
simple rule can say that it is not allowed to use the \textbf{log}
statement anymore.

The AOP methodology exists for different languages such as Java, .net
and C++. Here we have used AspectJ (see~\cite{AOPBible}) as one of the
oldest and best developed frameworks.

Using AOP instead of other technologies like wrappers
\cite{wrapper} has the advantage of  higher flexibility. Wrappers are
less powerful in this respect, as AOP offers a
higher modularisation.
This is in particular helpful in implementing security as
discussed in~\cite{AOPsec}. In~\cite{AOPsec2} the different problems
using container managed security are compared with a AOP based
approach.
\section{Role Based Access Control}\label{rbac}
The core component of any security system is the \textit{reference
  monitor} in which the security policy is defined.  This means that
the reference monitor mediates all accesses to \textit{objects} (methods, database
queries, service requests) by \textit{subjects} (users,
administrators, connected systems).

The present approach uses a modification of the Role Based Access Control (RBAC) 
model for the effective user rights.  A user joins one or more
roles and gets the combined access rights of the groups his or her
roles belong to.
A role describes an organisational work unit which changes only slowly
over the time. The concept of a \textit{workflow} is introduced to
describe the steps involved in a single user task.  In a
Web-based system a step is the transition form one page generated by
the Web-server to another.  In this view, \textit{Roles} are
collections of allowed workflows describing the work which may be
done by a user acting in this role.  RBAC is a protection item
centric (\cite[p. 129]{Rbac-Eck}, see also~\cite{RBAC-sand}) as opposed to a subject centric
description of access rights.

\begin{figure}[htbp]
   \centering \ifpdf

   \resizebox{0.5\textwidth}{!}{\includegraphics{securityStructure.png}}
   \else
\resizebox{0.5\textwidth}{!}{\includegraphics{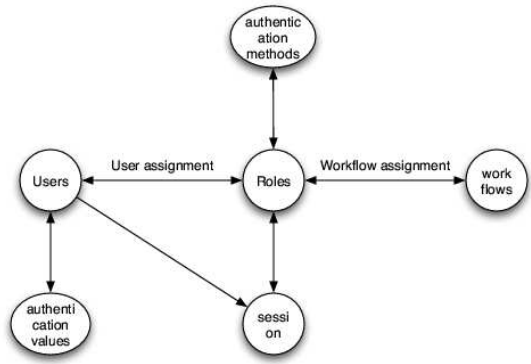}}
\fi
  \caption{RBAC structure}
  \label{fig:RBAC}
\end{figure}

Figure~\ref{fig:RBAC} shows the association between user, role, group,
and workflows.
This model ensures the possibility to define a security architecture
following the rule of the least privilege as described by Schroeder
and Saltzer~\cite{least}. The idea is that every subject only sees
items he or she is allowed to see or use.

As formalised by R. Sandhu and S. Kandala in~\cite{Workflow},
workflows and the RBAC model can be used in combination.  The access
rights are interpreted as the right to execute a workflow and are
called \textit{explicit}. This means that the user obtains access
rights through a finite state machine (FSM) computing the next
step in the process.  In this way the model grants \textit{implicit} access
rights which are changing over the time from page access to page
access. In effect, the user moves inside the application on a predefined path.
The realisation of workflows in a Web-based system needs a gentle way
to react if the user tries (wilfully) to leave the path. Most
graphical user interfaces have a common page the application is
starting with.  As a consequence every workflow contains at least this
page as a common state in the FSM definition, and fall-back to this
page is a commonly is a commonly used method in case of access right
violation attempts. It is also possible that two workflows have more than one page in
common, which is also to be addressed by an RBAC implementation.

Tools used supporting the administrator in the definition of security
rules must provide interfaces to define the RBAC model and the
workflows.
\section{Architecture}\label{arch}
\begin{figure}[htbp]
  \centering \ifpdf
  \resizebox{0.5\textwidth}{!}{\includegraphics{komp1.png}}
\else
  \resizebox{0.5\textwidth}{!}{\includegraphics{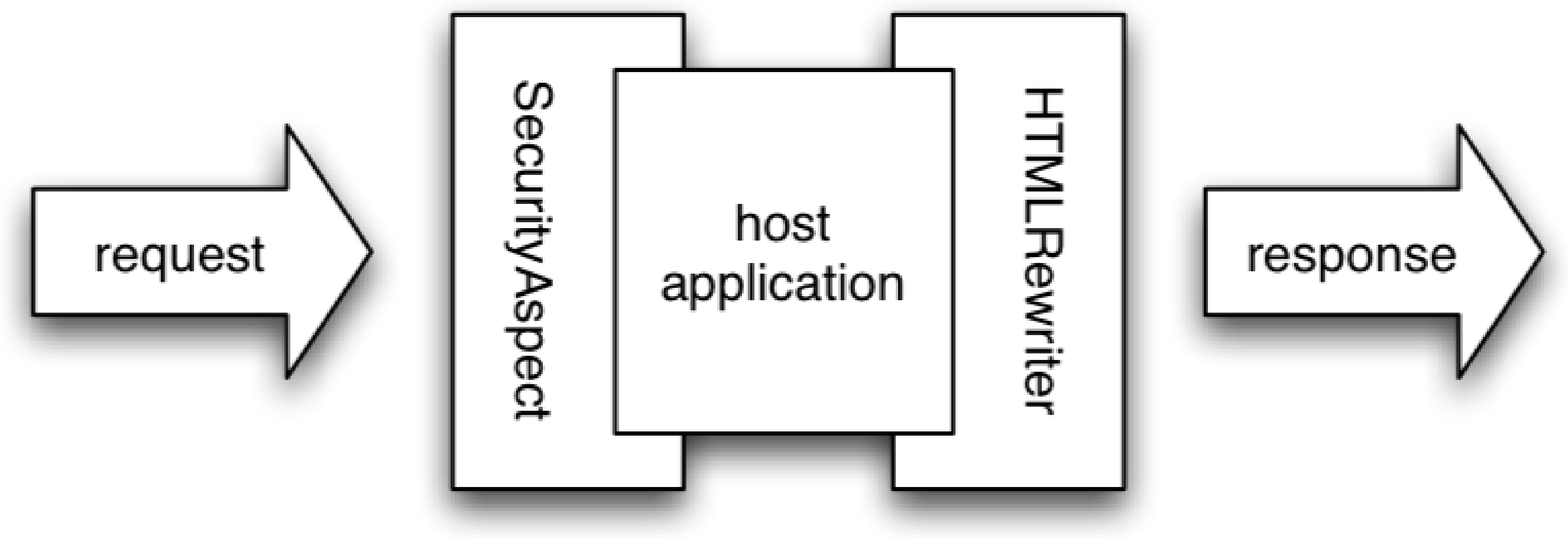}}
\fi
  \caption{Encapsulation of the host system}
  \label{fig:encap}
\end{figure}

Our security architecture encapsulates a Web-based system as shown in
Figure~\ref{fig:encap}. All incoming data is first processed by the
{\bf securityAspect} where the authentication and authorisation takes
place. If the user is in at least one allowed workflow for his role
the input is passed to the wrapped host system processing the
resulting Web-pages as usual. These pages can also be passed through
the {\bf HTMLrewriter} where checks and modifications in the html code
can be performed.
Capturing the in- and output of the system is, in a Web-based
application, easily implementable due to clear interfaces. Capturing
all input means to wrap two interface methods: {\bf get} and {\bf
  post} (in the case of the Tomcat container). As every result of the
host system is a direct result of the request by the user, this is
sufficient for authorisation control.  It is also thinkable to wrap
every method in the host system, thus enabling direct control over the 
its internal execution. It is then possible to exert extensive module testing, 
which can be useful for instance to test for buffer overflows or to find unusual patterns
in the program execution. This line of thought, though interesting,
was not further persecuted.

\begin{figure}[htbp]
  \centering \ifpdf
  \resizebox{0.6\textwidth}{!}{\includegraphics{komp2.png}}
  \else
  \resizebox{0.6\textwidth}{!}{\includegraphics{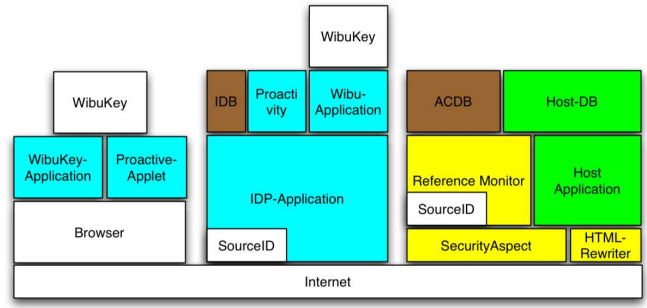}}
\fi
  \caption{Modules of the aspect-oriented security architecture.}
  \label{fig:Architektur2}
\end{figure}

In Figure~\ref{fig:Architektur2} the required modules are shown. 
Communication between them proceeds in vertical
direction. The host application which is extended in its
functionality is labelled green. The left block shows the software
modules running on the client and the middle block represents the IDP
modules. The hardware-token based authentication and
proactive re-authentication is detailed in Section~\ref{proactivity}.

The important task of local application login, i.e., 
mapping of a user from the federated identity to an
existing user known to the host system is, in the present
approach, solved by calling the host application's login page
from the securityAspect, and internally providing
password and user name to it.

The reference monitor has no knowledge of the host application at all
and decisions are made exclusively based on user input. 
Sometimes it is preferable to base a decision on the
state of the host application, to implement for instance, a rule that
can say that goods not in stock are not allowed to be sold. This
problem is approached by accessing the host database (see below).

Due to the task separation between IDP and SP (see Section~\ref{liberty})
two separate databases are required. The IDB stores all data
authenticating a single user (such as passwords).  The ACDB contains
the descriptions of workflows, the RBAC model and consequently also
the authentication methods required by the respective role
descriptions. The dependencies are illustrated in Figure~\ref{fig:RBAC}. 
Moreover the ACDB is used to store the states of the
active workflows for every session.  By this a user can have more than
one active session. 
In such an extension, the user would for example be able
to see different  user records at the same time. If this
side effect is not intended the Chinese Wall model~\cite{chinese} 
is an appropriate way
to describe these exclusions without completely denying the usage of parallel
sessions.
By refinements of the present security model it is
possible to define exclusive states or workflows implementing a
variation of the Chinese wall model.

As mentioned above the administrator of the security rules needs tools
for configuration of the securityAspect. This means
support for the  addition and specification of roles, users, workflows, and the
connections between them. Handling users and roles is a quite common
task.  Generating workflows is much more sophisticated as the
administrator has no knowledge of the system's source code. In view of
this we chose a training based approach. Training means here that the
actual system is used to define every required workflow by example.
To this end, the host system is weaved with a special training aspect
enabling the recording of user interaction. This aspect can be remotely
controlled by a GUI which in particular starts and stops the
recording. During recording every parameter returned by the
Web-site is shown to the trainer. The administrator can define 
admissibility rules for every
parameter using regular expressions \cite{regex}.
An advanced option is to define a SQL term describing a set including
the value of the parameter (see Figure~\ref{fig:workflow_editor}). As
basically every response of a Web-application is based on a
database state we believe that accessing the host database 
should solve most problems arising from the need to base access
control decisions on internal states of the host system.  The result
of the training phase is a single XML-file containing all data
required to construct the databases.

In the production phase, the host system is weaved with the
securityAspect. At run-time,
for every incoming request the securityAspect first checks if there was an
initial authentication of the corresponding user and if 
re-authentication is required. 
In case an initial authentication is needed, it
calls the liberty framework and the IDP to execute it. After
authentication, the securityAspect calls the security reference monitor
with all parameters of the request. The monitor then decides whether
the request is allowed or not, by determining the actual active
workflows and for every workflow its current state. It computes
the set of successor states and whether they can be reached with the
given parameters, taking into account the host database state. 
If the set of successor states is empty, the workflow is
marked as `not-active'. If all workflows are not active there is no
following state und hence the access is denied and an error page is
shown. From this error page the user can decide go back and retry
using different parameters or go to a predefined page, which could, e.g., be
the initial page of at least one workflow of his/her role. 
As stated above all workflows generically have a start page in common which is the
page preferably used as base page in an error situation. After the
host system has processed the input, it returns an HTML page to the
client. Here, the HTMLrewriter rewrites the Web-page and integrates the
applet which provides the proactive functionality.

As a side benefit, collecting audit data is easy in this model. Every
action can be recorded and stored in a database in a personalised way.
Actions are more than database accesses, in particular, every
interaction with the host system can be logged, so it is in theory
possible to recognise common flaws and mistakes in the user
interaction or users with long idle times. Of course this may raise
privacy concerns.
\section{Deployment Scenario}\label{proto}
In this section we give a short overview of the deployment process of
the prototypical implementation previously described.

\begin{figure}[htbp]
  \centering \ifpdf

  \resizebox{0.5\textwidth}{!}{\includegraphics{workflow-generator2.png}}
  \else
  \resizebox{0.5\textwidth}{!}{\includegraphics{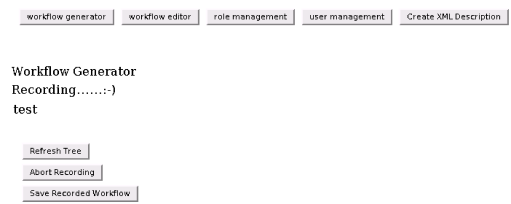}}
\fi
  \caption{Workflow recording.}
  \label{fig:workflow_recording}
\end{figure}

\begin{figure}[htbp]
  \centering\ifpdf
  \resizebox{0.5\textwidth}{!}{\includegraphics{workflow-editor3.png}}
  \else
  \resizebox{0.5\textwidth}{!}{\includegraphics{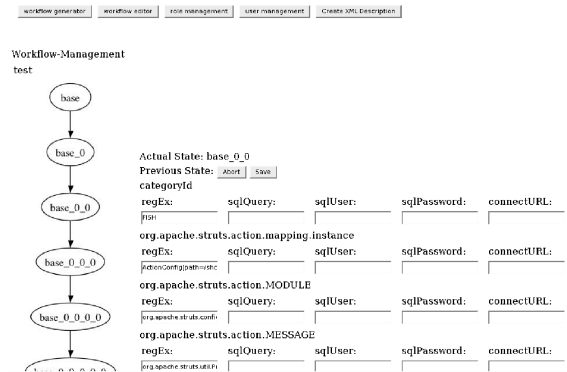}}
\fi
  \caption{Workflow editor.}
  \label{fig:workflow_editor}
\end{figure}

Figure~\ref{fig:workflow_recording} shows a screen-shot of a workflow
recording with the webtool, the main utility for the training phase.
The recording has already been started and concurrently running
the application the administrator is able to create a workflow description
by simply clicking through it.
As we are using a `minimal need to know' security strategy~\cite{leastPriv} 
it is not necessary to define every possible
exceptional state as every state not defined in the workflow is
forbidden. Here, a gentle reaction of the security module is required
supporting the user returning back into his workflow.
The approach hides many complexities of, for instance, collaborative
workflows involving more than one user.
After saving the workflow it is possible (see Figure~\ref{fig:workflow_editor}) 
to edit the recorded workflows.
In particular, one can change the parameters of
a submitted page in order to verify them in the production phase
against a SQL database query or regular expressions.

\begin{figure}[htbp]
  \centering \ifpdf
  \resizebox{0.5\textwidth}{!}{\includegraphics{user-editor3.png}}
  \else
  \resizebox{0.5\textwidth}{!}{\includegraphics{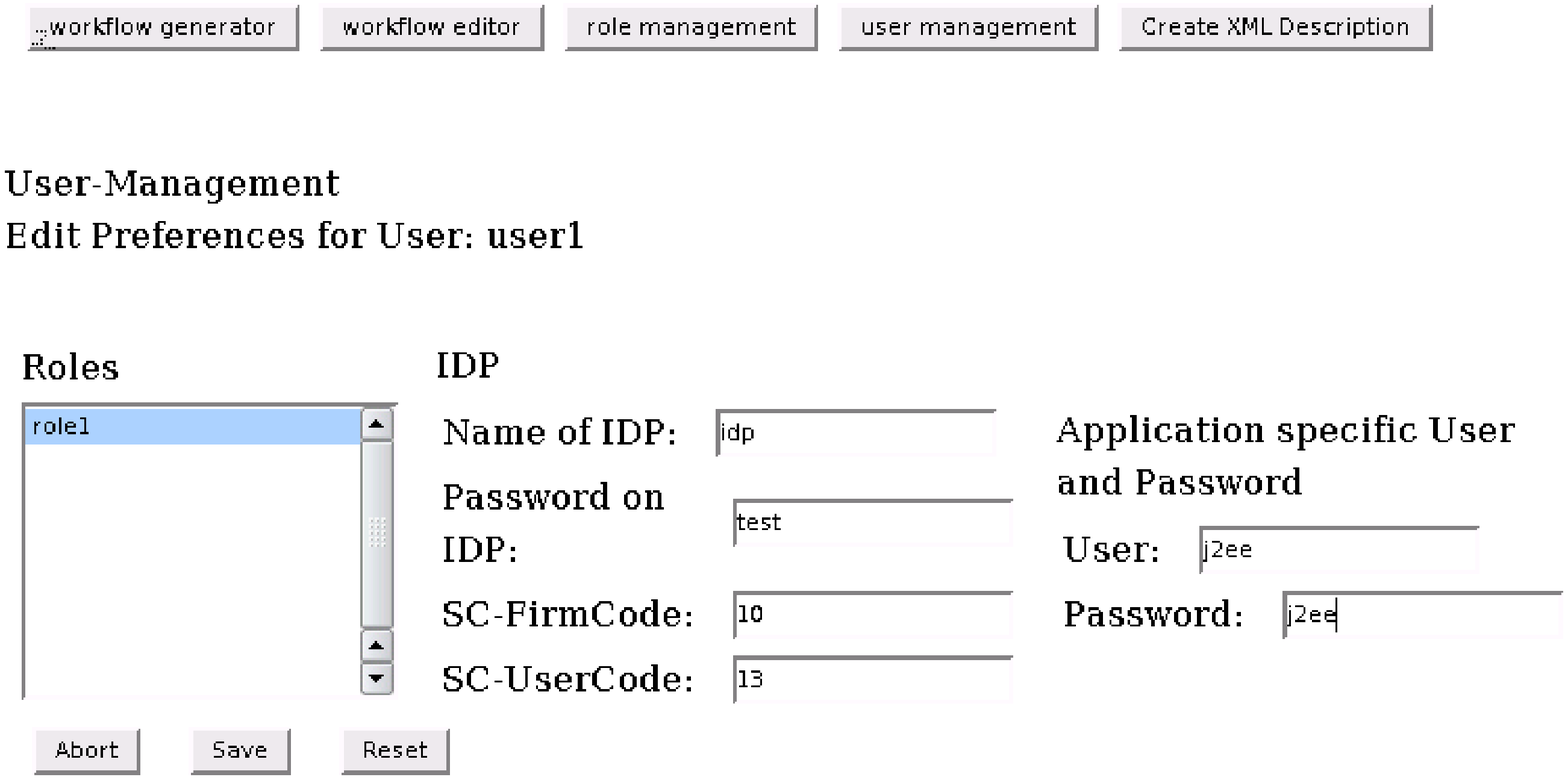}}
\fi
  \caption{User editor.}
  \label{fig:user_editor}
\end{figure}

When all required workflows have been recorded, roles and users are
created as shown in Figure~\ref{fig:user_editor}.  Every user has to
be a member of at least one role and has to be assigned to one IDP.
For the authentication by the IDP, every user gets a password and if
he uses the Wibu-Key hardware token a Wibu-Key specific FirmCode and
UserCode is assigned to identify the specific token.  If the user is to be
authenticated against the application user management system, the
application specific user name and password must be provided, too.

\begin{figure}[htbp]
  \centering \ifpdf
  \resizebox{0.5\textwidth}{!}{\includegraphics{login2.png}}
  \else
  \resizebox{0.5\textwidth}{!}{\includegraphics{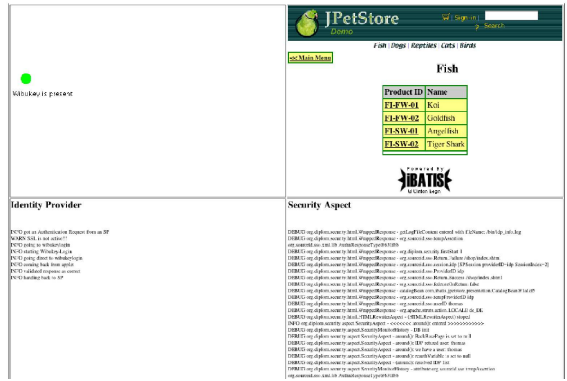}}
\fi
  \caption{Productive phase}
  \label{fig:productive_phase}
\end{figure}

At the beginning of the production phase, the securityAspect is weaved
with the host application. With the HTMLRewriter aspect we changed the
output of the application and added different debug messages and the
applets for the Wibukey (see Figure~\ref{fig:productive_phase}). The
green circle of the Wibukey applet shows that the Wibukey is attached
to the PC and the user has shown activity in the observed period of time. If
the hardware token is attached to the client whithout ongoing
interaction, the light turns to red.
\section{Conclusion}\label{con}
We have proved that it is possible to add AAA capabilities to an
existing Web-based application without any knowledge of the source code of the
host system. Furthermore we are
able to `plug-in' a workflow-based security policy which can be
adapted in an intuitive way.
By this we have also shown that aspect oriented programming is an
appropriate tool for implementing security features into existing and
new systems. It opens new possibilities in modularisation and code
re-usage for system security.

An SSO enabled system offers a new quality in business interaction.
Using the liberty framework it is possible to establish trust
relationships with very little effort. In combination with 
workflows and the training suite, also small companies are
enabled to interact with others and to establish Web-enabled
workflows, e.g., for field personnel.

From a security perspective, the weak point of the present architecture
is the single IDP. A denial-of-service attack can be used to make it
impossible to use any service depending on the federated identity.
This means that the IDP must be protected and that there is a need for
a fall-back strategy. A second threat is a disclosure of the account
data stored in the ACDB database.

Another problem lies in the high modularisation of security. 
As there is no direct connection between security and host
system it is not possible to define rules allowing for consideration
of internal states of the host system. We approached this issue by
introducing SQL terms operating on the host database.

On the other hand a high modularisation means that it is possible to
develop a `plug-in' security module once and to reuse it in
different products. This reduces costs and facilitates testing, 
thereby easing security evaluations. If there is a
failure in the security module it is very easy to replace this part
and to insert the optimised version into the system even at production
time.
\gdef\etalchar#1{$^{#1}$}


\begin{thebibliography}{99}

\bibitem{Khalf04} 
Abdulwahed Mo. Khalfan.
Information security considerations in IS/IT outsourcing projects: 
a descriptive case study of two sectors.
International Journal of Information Management
\textbf{24} (2004) 29--42.

\bibitem{DACH04-Mehrhoff} 
Michael Mehrhoff.
\newblock Outsourcing unter Sicherheitsaspekten.
\newblock In: D.A.CH Security 2004, P.Horster (Ed.),
syssec, 2004,  pp.\ 62--70

\bibitem{WBEM} 
Jeff Prosise.
\newblock DMTF Web-Based Enterprise Management Initiative.
\\ \newblock URL: \url{http://www.dmtf.org/standards/wbem}

\bibitem{DACH04-MakoSi}
Michael Herfert, Andreas~U. Schmidt, Peter Ochsenschläger, Jürgen Repp,
  Roland Rieke, Martin Schmucker, Steven Vettermann, Uwe Böttge, Cristina
  Escaleira und Dirk Rüdiger.
\newblock Implementierung von Security Policies in offenen Telekollaborationen.
\newblock In: D.A.CH Security 2004, P.Horster (Ed.),
syssec,  2004, pp.\ 37--49.

\bibitem{AOPBible}
Ramnivas Laddad.
\newblock AspectJ in Action: Practical Aspect-Oriented Programming.
\newblock Manning Publications Co., 2003.

\bibitem{libertyTutorial} 
Liberty~Alliance Project.
\newblock Liberty Alliance Developer Tutorial.
\\ \newblock URL: \url{http://www.projectliberty.org/resources/tutorial_draft.pdf}.

\bibitem{RBAC} 
D.F. Ferraiolo und D.R. Kuhn.
\newblock Role Based Access Control.
\newblock 15th National Computer Security Conference, 1992.
\newblock URL:
  \url{http://csrc.nist.gov/rbac/Role\_Based\_Access\_Control-1992.html}.

\bibitem{wibukey} 
Wibukey~Systems AG.
\newblock WIBU-KEY.
\newblock URL: \url{http://wibu.de/de/wibukey.php}.

\bibitem{IndetityPrimer}
Ping Identity Corporation
\newblock Federated Identity Primer
\newblock URL: \url{www.pingidentity.com}

\bibitem{liberty}
Liberty Alliance
\newblock URL: \url{http://project-liberty.org}.

\bibitem{SourceID} 
SourceID - Open Source Federated Identity Management
\newblock URL: \url{http://www.sourceid.org}.

\bibitem{SAML}
Liberty Alliance Project
\newblock Liberty Alliance Extends Liberty Interoperable Testing Program to Include SAML 2.0 OASIS Standard
\newblock URL: \url{https://www.projectliberty.org/press/details.php?item_id=111}

\bibitem{SAML20}
OASIS
\newblock OASIS Security Services
\\ \newblock URL: \url{http://www.oasis-open.org/committees/tc_home.php?wg_abbrev=security\#samlv20}

\bibitem{proactive} 
Antti Salovaara and Antti Oulasvirta.
\newblock Six modes of proactive resource management: a user-centric typology for proactive behaviors.
\newblock Proceedings of the third Nordic conference on Human-computer interaction,
\newblock ACM Press, 2004, pp.\ 57--60
\newblock URL: \url{http://doi.acm.org/10.1145/1028014.1028022}.

\bibitem{AOP_start1}
Gregor Kiczales, John Lamping, Anurag Mendhekar, Chris Maeda, Cristina Videira Lopes,
Jean-Marc Loingtier und John Irwin.
\newblock Aspect-Oriented Programming. 
\newblock 1997. 
\\ URL: \url{http://www2.parc.com/csl/groups/sda/publications/papers/Kiczales-ECOOP97/}

\bibitem{AOP_start2}
Anurag Mendhekar, Gregor Kiczales und John Lamping. 
\newblock RG:A Case-Study for Aspect-Oriented Programming. 1997. 
\newblock URL: \url{http://www2.parc.com/csl/groups/sda/publications/papers/PARC-AOP-RG97/}

\bibitem{AOSD}
Bart De Win, Wouter Joosen und Frank Piessens. 
\newblock AOSD \& Security: a practical assessment.
\newblock February 28, 2003
URL: \url{http://www.cs.kuleuven.ac.be/cwis/research/distrinet/resources/publications/41067.pdf}

\bibitem{wrapper}
Thomas Bühren, Volker Gruhn und Dirk Peters. 
\newblock Konfigurierbare Sicherheit für Java Laufzeitumgebungen. 
\newblock D.A.CH Security, 2004, pages 329--340

\bibitem{AOPsec}
Bart De Win, Wouter Joosen, and Frank Piessens. 
\newblock AOSD as an enabler for good enough security.
\newblock URL: \url{http://www.cs.kuleuven.ac.be/cwis/research/distrinet/resources/publications/41066.pdf}

\bibitem{AOPsec2}
Pawel Slowikowski and Krzysztof Zielinski
\newblock Comparison Study of Aspect-oriented and Container Managed Security
\newblock In: AAOS2003: Analysis of Aspect Oriented Software, 2003.

\bibitem{Rbac-Eck}
C. Eckert.
\newblock IT-Sicherheit.
\newblock Oldenbourg Wissenschaftsverlag GmbH
\newblock 2001

\bibitem{RBAC-sand}
R. Sandhu. 
\newblock Role-based access control. In Advances in Computers, 
\newblock Vol.\ 46, pp.\ 237--286. M. Zelkowitz Eds. Academic, 1998

\bibitem{least}
Jerome H. Saltzer und Michael D. Schroeder 
\newblock The Protection of Information in Computer Systems. 1975
\newblock URL: \url{http://web.mit.edu/Saltzer/www/publications/protection/}

\bibitem{Workflow}
Savith Kandala und Ravi Sandhu. 
\newblock Secure role-based workflow models. 
\newblock In: Proceedings of the fifteenth annual working conference on Database and application security, 
\newblock Kluwer Academic Publishers, 2002, pp.\ 45--58

\bibitem{chinese}
D. F. C. Brewer und M. J. Nash
\newblock The Chinese Wall Security Policy
\newblock In: IEEE Symposium on Security and Privacy, 1989, pp.~206--214

\bibitem{regex}
Jeffrey E. F. Friedl.
\newblock Mastering Regular Expressions. 
\newblock O'Reilley \& Associates, Cambridge, MA, 1997 

\bibitem{leastPriv}
Fred B. Schneider.
\newblock Least Privilege and More. 
\newblock j-IEEE-SEC-PRIV, 1(5):55--59, September/October 2003. 
\newblock URL: \url{http://csdl.computer.org/dl/mags/sp/2003/05/j5055.pdf}
%
\end{thebibliography}
\end{document}